\documentclass[a4paper,fleqn,usenatbib]{mnras}

\usepackage{newtxtext,newtxmath}

\usepackage[T1]{fontenc}
\usepackage{ae,aecompl}

\usepackage{graphicx}	
\usepackage{amsmath}	
\usepackage{amssymb}	
\usepackage{subfigure}
\usepackage{placeins}

\title[]{Does the SN rate explain the very high energy cosmic-rays in the central 200pc of our Galaxy?}

\author[L. Jouvin et al.]{
L. Jouvin,$^{1}$\thanks{E-mail: lea.jouvin@apc.in2p3.fr}
A. Lemi\`ere,$^{1}$
and R. Terrier $^{1}$
\\
$^{1}$APC, AstroParticule et Cosmologie, Universit\'e Paris Diderot, CNRS/IN2P3, CEA/Irfu, \\ Observatoire de Paris, Sorbonne Paris Cit\'e, 10, rue Alice Domon et L\'eonie Duquet, 75205 Paris Cedex 13, France.
}

\date{Accepted XXX. Received YYY; in original form ZZZ}

\pubyear{2015}

\begin{document}
\label{firstpage}
\pagerange{\pageref{firstpage}--\pageref{lastpage}}
\maketitle


\begin{abstract}
The H.E.S.S. collaboration revealed the presence of a very high energy (VHE) diffuse emission in the inner 100pc of the Galaxy in close correlation with the Central Molecular Zone (CMZ). Recently, they deduced from this emission a cosmic-ray (CR) over-density in the region with a local peak toward the Galactic Center (GC) and proposed a diffusive scenario with a stationary source at the GC to explain it. However, the high supernovae (SN) rate in the GC must also create a sustained CR injection in the region via the shocks produced at the time of their explosion. Considering typical diffusion coefficient close the interstellar medium (ISM) value yields to a diffuse escape time much lower than the recurrence time between each SN explosion, showing that a steady state model will fail to reproduce the data. This work aims to study the impact of the spatial and temporal distribution of SNs in the CMZ on the VHE emission morphology and spectrum: we build a 3D model of VHE CR injection and diffusive propagation with a realistic gas distribution. We show that a peaked $\gamma$-ray profile towards the GC can be obtained using realistic SN spatial distribution taking into account the central massive star cluster. We conclude that the contribution of SNs can not be neglected, in particular at large longitudes, however an additional CR injection at the GC is required to reproduce the very central excess. 
\end{abstract}



\section{Introduction}
\label{intro}
An excess of CRs was discovered in the center part of our Galaxy with the detection of a hard emission of very high energy $\gamma$-ray (100 GeV-100 TeV) by the High Energy Stereoscopic System (H.E.S.S.). This diffuse emission is localized along the Galactic plane and is extending over 2$^\circ$ in longitude and 0.3$^\circ$ in latitude \citep{2006Natur.439..695A}. This particular region of the Galaxy contains 1.2-6.4$\, \times 10^7$ M$\rm_{\odot}$ of interstellar $H_2$ gas \citep{1998A&A...331..959D}, $\sim$ 10\% of the total molecular mass of the Galaxy, in a rather complex setup of dense molecular clouds (density $\sim 10^4$ $\rm cm^{-3}$) called the Central Molecular Zone (CMZ). The CMZ extends to 300 pc in Galactic longitude and 100 pc in latitude. The close correlation between the VHE $\gamma$-ray emission and the target material suggests that the dominant component of the ridge is due to the interaction of relativistic cosmic-rays (CRs) with protons of the ambient medium. The CR density deduced from this emission is between 3 and 9 times higher than the one measured on Earth and with a harder spectrum, $\Gamma\sim \, $2.3. Recently, \citet{2016Natur.531..476H} confirmed the presence of this extended VHE emission by analysing 10 years of H.E.S.S. data. In particular, the CR profile they deduced from the VHE emission is peaked towards the Galactic Center (GC) and compatible with a 1/r profile as the one expected from a stationary pointlike injection of CR located at the Galactic Center.

The center of the Galaxy hosts a super-massive black hole (SMBH), SgrA$^\star$, of about $4 \cdot 10^6$ M$_{\odot}$ \citep{2008ApJ...689.1044G}. Nowadays, this SMBH is extremely inactive and is fed by the stellar winds in this region. The resulting hot accretion flows are radiatively inefficient and are most likely associated with strong outflows \citep{2014ARA&A..52..529Y, 2013Sci...341..981W} where the CR acceleration could take place \citep{2006ApJ...647.1099L}. With an actual bolometric luminosity of $10^{36} \, \rm{erg\,s^{-1}}$ , most of the accretion power in the Bondi region, $10^{39} \, \rm{erg \cdot s^{-1}}$ \citep{2013Sci...341..981W}, could potentially be used to accelerate CR up to PeV energies. \citet{2016Natur.531..476H} detailed the possible relation between the VHE emission and a stationary central source, as SgrA$^\star$.

The environment in the GC is often compared with the one of a starburst system. Around 2\% of the Galaxy's massive star formation occur in this region. Many Supernova Remnants (SNR) are visible through their radio and X-ray thermal emission \citep{2015MNRAS.453..172P}, and Pulsar Wind  Nebulae (PWN) are also numerous in this rich part of the Galaxy \citep{2007yCat..21650173M, johnsonJ}. Extrapolating the Galactic SN rate to the GC star formation rate, \citet{2011MNRAS.413..763C} expect 0.04 SN per century in the central 300 pc. However this rate suffers from large uncertainties. The range determined from stellar composition by \citet{2011MNRAS.413..763C} goes from 0.02 to 0.08 SN per century. Counting the number of X-ray bright SNR candidates in the GC, \citet{2015MNRAS.453..172P} estimated a SN rate, $r_{SN} \sim 0.035-0.15 \, \mathrm {per \, century}$, compatible with the previous estimation. This high SN rate in the region must create a sustained CR injection via the shocks produced at the time of their explosion. Knowing that the kinetic energy released from a SN explosion is $\rm 10^{51} \, erg$ and assuming a SN rate of $\rm 4 \times 10^{-4} \, {yrs}^{-1}$, the injection power in the GC from these sources is around $\rm 10^{40} \, erg \cdot s^{-1}$. The $\gamma$-ray luminosity observed by H.E.S.S. is several order of magnitude lower than this input power \citep{2011MNRAS.413..763C}. Therefore, any model of $\gamma$-ray emission in the region has to take into account the propagation and the escape of CRs injected by these multiple accelerators.

The GC harbors three compact and massive star clusters: the Quintuplet, the Arches and the Central disk surrounding SgrA$^\star$. Around 1/3 of the massive stars detected in the GC are located outside of these three massive starburst clusters suggesting the existence of isolated high mass star formation \citep{2010ApJ...725..188M}. However these isolated stars could have been kicked out from the Arches or the Quintuplet cluster due to their dynamics \citep{2014A&A...566A...6H}. The presence of these young clusters at the center indicates that even if the real SN spatial distribution is unknown, it is clearly non uniform and concentrated in the inner 30 pc. Therefore it could also create a CR density peaking towards the GC. 

This work aims to study the impact of CR accelerated by the SNRs on the VHE emission observed in the inner 200 pc of the Galaxy. We want to emphasize that any scenario of CR injection at the center \citep{2015MNRAS.451.1833M, 2016Natur.531..476H} cannot be studied separately from these SNR contributions. Following the work of \citet{2014ApJ...790...86Y} or \citet{2015MNRAS.451.1833M}, we first model the overall emission due to these sources with a one zone stationary steady state scenario in section \ref{section1}. We compare the case of an advective or a diffusive CR escape from the GC considering typical diffusion coefficient values in the interstellar medium (ISM). Because of the large diffusive escape time compared to the SN rate, we conclude that steady state cannot be assumed. We then explore the impact of the actual matter and SN spatial distribution by building a 3D model of CR and gas distribution in the section \ref{model}. Finally, in section \ref{results}, we present and discuss the results obtained with this 3D model. In particular, we study how the spatial distribution of these CR accelerators throughout the CMZ contributes to the CR excess seen by H.E.S.S. and compare it to a stationary source at the center such as, possibly, SgrA$^\star$.

\section{One zone steady state model: Advection vs Diffusion}
\label{section1}
In this section, the CR population arising from the multiple SNe in the GC is produced using a one zone steady state model following the work of \citet{2014ApJ...790...86Y,2011MNRAS.413..763C}. In these studies, the presence of a perpendicular high speed wind, from 400 to 1000 $\rm km \, s^{-1}$, is required for the CRs to escape. Taking common value for the diffusion coefficient in the ISM, we argue that a diffusive escape is dominant over the advective escape. In addition, the observed $\gamma$-ray luminosity is compatible with the high SN rate in the region in a diffusive scenario. In this case, the non stationary effects cannot be neglected.

\subsection{Advection: energy independent escape}
\label{advection}
We consider a steady state CR injection in a box. The CRs escape from the box due to the advection of a perpendicular wind of speed $v$ on a length scale $H$ ($\tau_{adv}=H/v$). We consider a typical scale for CRs to escape of $H=50 \, \rm{pc}$. In a steady state scenario:
\begin{center}
\begin{eqnarray}
Q-\frac{N}{\tau_{adv}}=0
\label{steadystate}
\end{eqnarray}
\end{center} 
where $N$ is the total number of CRs per energy unit ($\rm{TeV^{-1}}$) and $Q$ the spectrum injected per unit of time assumed to be distributed as a power-law, $Q=Q_0\times$E$^{-p}$. The normalisation $Q_0$ of the spectrum is given by the SN injection rate assuming that they accelerate CR between 1 GeV and 1 PeV:

\begin{center}
\begin{eqnarray*}
\int_{1 \, GeV}^{1 \, PeV} Q_0\times E^{-p} \, dE = \eta \, E_k \, r_{SN}
\label{steadystate2}
\end{eqnarray*}
\end{center} 
where $E_{k}$ is the total energy released at the SN explosion, $\eta$ the efficiency of CR acceleration and $r_{SN}$ the SN rate.

The VHE $\gamma$-ray emission observed with H.E.S.S. presents a hard spectrum with an index $\Gamma$ of 2.3 \citep{2006Natur.439..695A, 2016Natur.531..476H}. The advection being an energy independent process, it doesn't modify the spectral index. Taking into account the spectral hardening due to pion production and decay \citep{2014PhRvD..90l3014K}, the data are well reproduced by a power law injection of index $p$ equal to 2.45. Considering this spectral index, we deduce from the relation above:

\begin{center}
\begin{eqnarray*}
Q_0=4\times 10^{36} \, \rm{TeV}^{-1} s^{-1} \left(\frac{\eta}{10\%}\right) \left(\frac{E_{k}}{10^{51}\, erg}\right) {\left(\frac{r_{SN}}{10^{-4} \, yrs^{-1}}\right)}
\label{steadystate2}
\end{eqnarray*}
\end{center}

The $\gamma$-rays luminosity produced by these CRs , $L_{\gamma}(>200 \, \rm{GeV})$, is related to the total energy of CRs ($\rm{erg}$), $W_{p}(>2 \, \rm{TeV})$=$\int_{>2 \, \rm{TeV}} N E dE$, by the following relation:

\begin{center}
\begin{eqnarray*}
L_{\gamma}(>200 \, \rm{GeV}) \sim \frac{W_{p}(>2 \, \rm{TeV})}{\tau_{pp->\pi_0}} \\
\end{eqnarray*}
\end{center}

where $\tau_{pp->\pi_0}$=$4.4 \times 10^{15} \left( \frac{n}{1 \, \rm{cm^{-3}}}\right)^{-1} \rm sec$ is the proton energy loss time scale due to neutral pion production in an environment of hydrogen gas of density $n$ \citep{2004vhec.book.....A}. Hence we have:

\begin{center}
\begin{multline}
L_{\gamma}=3.4\times 10^{35} \, \rm{erg \cdot s^{-1}} \left(\frac{\eta}{10\%}\right) \left(\frac{E_{k}}{10^{51} \rm{erg}}\right) {\left(\frac{r_{SN}}{10^{-4} \, \rm{yrs^{-1}}}\right)} \left(\frac{H}{50 \, \rm{pc}}\right) \\ \times {\left(\frac{v}{10^{3} \, \rm{km/s}}\right)}^{-1} \left(\frac{n}{100 \, \rm{cm^{-3}}}\right)
\label{steadystate4}
\end{multline}
\end{center} 

in good agreement with the measured $\gamma$-ray luminosity reported by \citet{2006Natur.439..695A}, L$_\gamma$=$3.5\times 10^{35} \, \rm erg \cdot s^{-1}$. Here we take a typical value for the gas density $n$ of $100 \, \rm{cm^{-3}}$ as expected considering typical size and total mass estimation of the CMZ \citep{2007A&A...467..611F} as discussed above. Regarding the uncertainties of the total mass estimations of the CMZ, this density can vary from a factor 2 to 4.

Even by considering a very large wind speed of 1000 km/s, the recurrence time between each SN explosion $\tau_{SN}$ has to be around $10\rm^4 \, years$ ($\rm{r_{SN}}=$1/$\tau_{SN}=10^{-4} \, yrs^{-1}$), in order not to overproduce the total $\gamma$-ray luminosity observed by H.E.S.S.. This value is high compare to several estimations by \citet{2014ApJ...790...86Y, 2011MNRAS.413..763C}. However some studies estimate a rather low star formation rate (SFR) in the CMZ \citep{2015ApJ...799...53K, 2013MNRAS.429..987L} leading to a SN rate compatible with the previous value of $\rm{r_{SN}}=10^{-4} \, \rm{yrs^{-1}}$ assuming a standard IMF \citep{2001MNRAS.322..231K}. However, the IMF in the GC may be different from the standard one. With a more top heavy IMF, the SN rate deduced from the SFR could be higher. In particular, using the observed IMF and the total mass estimated of the central part of the massive star clusters, the Quintuplet and the Central cluster, we already find a SN rate around $0.3-1.2 \times 10^{-4} \, \mathrm {yrs}^{-1}$ (see section \ref{CR_accelerators}). Therefore, based on the rate already found in each of the two massive star clusters of the region, a SN rate of $\rm{r_{SN}}=10^{-4} \, yrs^{-1}$ in the GC seems really low but it belongs to the lower edge of SN rate estimations.

Moreover, to reproduce the spectral index of the $\gamma$-ray spectrum \citep{2006Natur.439..695A}, we have to consider an injection spectrum with an index $p$ equal to 2.45 which is rather soft for particle acceleration at SNR shock. The CR escape is not sufficiently fast when considering the advection only. 

In the section \ref{diffusion}, we discuss the case of a diffusive escape which is dominant over an advective escape if considering a common diffusion coefficient value. 

\subsection{Diffusion: energy dependent escape} 
\label{diffusion}
When modelling an energy dependent escape by diffusion, the value chosen for the diffusion coefficient is critical. For the GC, its value is unknown. Several studies \citep[e.g.][]{2008MNRAS.387..987W} tried to estimate its value numerically by simulating particle trajectory in a totally turbulent magnetic field through which the hadrons must diffuse. As expected for strong turbulent magnetic field, the coefficient derived is very small and excludes the possibility of a diffusive escape. Indeed at these low values the advection is more competitive than the diffusion \citep{2002PhRvD..65b3002C}. This is why they ruled out the possibility of localized source accelerators like the central black hole or the SNe along the disk and concluded with the necessity of diffuse stochastic particle acceleration.

However, several independent observations reveal the presence of a relatively strong and ordered magnetic field throughout the CMZ \citep{2011IAUS..271..170F, 2014arXiv1406.7859M}. In the intercloud medium, on average the magnetic field is poloidal. Indeed, numerous non thermal radio filaments (NRFs) have been detected, most of them oriented perpendicularly to the Galactic plane. The strong polarization of their synchrotron emission indicates that the magnetic field points along the filaments. This large scale structured magnetic field in the GC indicates the possibility of higher diffusion coefficient values \citep{2002PhRvD..65b3002C}. 

In this section, we assume a power-law diffusion coefficient: $D=D_0 {\left(E/10 \ \rm{TeV}\right)}^{\rm d}$. The dependence $d$ of the diffusion coefficient depends on the power spectrum of turbulence of the $B$ field. We assume a Kolmogorov spectrum which gives a diffusion coefficient index $d=1/3$.

The diffusion time is given by $\tau_{diff}=H^2/(4 \times D)$. A common value of the diffusion coefficient at 10 GeV in the Galactic ISM is $D_{10 \, \rm{GeV}} \sim 10^{28}$ $\rm cm^2s^{-1}$ \citep{1990acr..book.....B}. For this usual diffusion coefficient and a proton at 1 TeV, the diffusion is already more competitive than the advection since $\tau_{diff} < \tau_{adv}$. In the following, we will normalize the diffusion coefficient at 10 GeV, $D_0$, in order to have a value of $2\times 10^{29}$ $\rm cm^2s^{-1}$ at 10 TeV close to ISM values \citep{2014PhRvD..89h3007G, 2011ApJ...729..106T} \footnote{We choose a conservative value for the diffusion coefficient at 10 TeV. Knowing the magnetic field in the GC is higher than in the general ISM, we expect a diffusion coefficient at the lower boundaries of the one found in \citet{2014PhRvD..89h3007G, 2011ApJ...729..106T} since it depends of the magnetic field strength \citep{1990acr..book.....B}.}. Since the escape is energy dependent for the diffusion, to reproduce the VHE $\gamma$-ray spectrum, the proton spectral index $p$ is fixed to 2.15. 
 
The $\gamma$-ray luminosity is given by:
\begin{center}
\begin{multline}
L_{\gamma} (10 \, \rm{TeV})=3.2\times 10^{35} \, \rm {erg} \, \rm {s^{-1}} \left(\frac{\eta}{10\%}\right) \left(\frac{E_{k}}{10^{51} \, \rm{erg}}\right) {\left(\frac{r_{SN}}{5\times 10^{-4} \, \rm{yrs}^{-1}}\right)} \\ \times {\left(\frac{H}{50 \, \rm{pc}}\right)}^2 
{\left(\frac{D_0}{2 \times 10^{29} \, \rm{cm^2s{-1}}} \right)}^{-1} \left(\frac{n}{100 \, \rm{cm^{-3}}}\right)
\label{steadystate5}
\end{multline}
\end{center}

Compared to the advective escape, the proton spectrum injected is harder and the power injected is more uniformly distributed over the proton spectrum. With the softer spectrum such as the one we have to assume in the advection scenario, most of the power is injected in the low energy part. Therefore it artificially minimizes the problem of the overproduction of the total $\gamma$-ray luminosity at the high energy part of the spectrum. The high energy protons escape faster from the box when considering this energy dependent escape time. As a result, the recurrence time for the SNe has to be lower than for the advective escape. To match the measured TeV flux, the rate has to be around $5\times 10^{-4} \, \rm{yrs}^{-1}$, in the typical range of the SN rate estimations in the GC (see section \ref{intro}).

\subsection{Spectral Energy Distribution for the advective and diffusive escape} 
We compare the multiwavelength spectral energy distribution (SED) for these one zone steady state models in the case of an advective escape (section \ref{advection}) or diffusive escape (section \ref{diffusion}) with existing radio, GeV and TeV data. In order to obtain these SEDs, we have to solve the kinetic equation:

\begin{eqnarray}
\frac{\delta N}{\delta t} = \frac{\delta }{\delta \gamma}(PN) - \frac{N}{\tau} + Q
\end{eqnarray}
  where $N$ is the spectral particle density, $\gamma$ is the particle Lorentz factor, $P$ the energy loss rate, $\tau$ the loss time-scale and $Q$ the source spectrum. 
  
We used the software package GAMERA \citep{gamera}. The energy loss processes taken into account are the ionization, synchrotron, bremsstrahlung, inverse compton (IC) and inelastic pp scattering. We assume a maximum acceleration energy of 1 PeV for the protons and 1 TeV for the electrons since radiative losses are expected to be already very strong for 1 TeV electrons. Both in the advective or diffusive scenari, the efficiency for CR acceleration is fixed to 10\% of the kinetic energy released from a SN explosion and the ratio electrons/protons is assumed to be 1\%. As considered in the previous sections \ref{advection} and \ref{diffusion}, the typical scale $H$ for CRs to escape is 50 pc. For the IC, we considered two target photon populations as \citet{2014ApJ...790...86Y}: an optical radiation field (T=5000K) with an energy density of 60 $\rm{eV \, cm^{-3}}$ and a far infrared radiation field (T=21 K) with an energy density of 15 $\rm{eV \, cm^{-3}}$. 

Outside the NRFs, the diffuse radio emission shows a downward break at 1.7 GHz that constrains B $\geqslant$ 50 $\mu G$ \citep{2011MNRAS.413..763C}. However, the magnetic field in the general ISM could be close to the equipartion with the cosmic-rays, $B_{eq}\approx 10$ $\mu G$ \citep{2005ApJ...626L..23L}. We use different magnetic field strengths between 10 and 100 $\mu G$. Whatever the value, the IC is never dominant for the GeV and TeV domain.

\begin{figure}
\centering
{\includegraphics[width=\columnwidth]{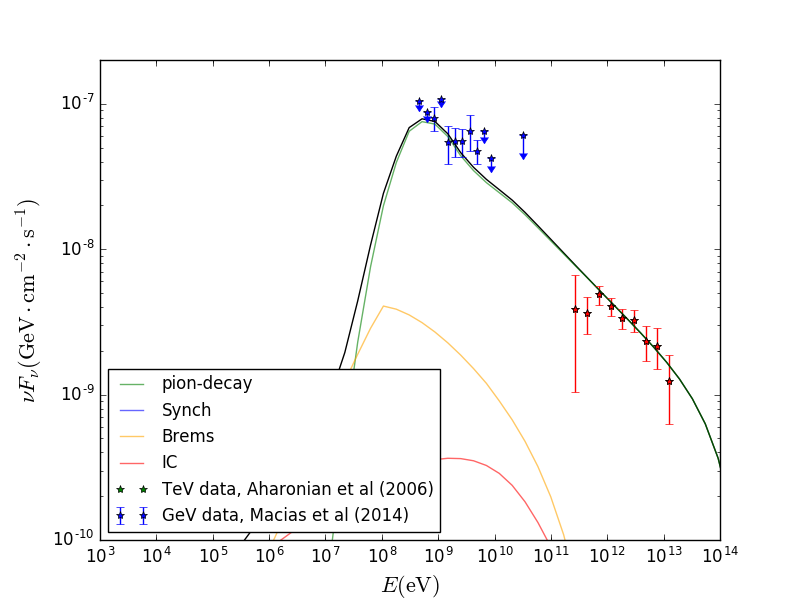}}
{\includegraphics[width=\columnwidth]{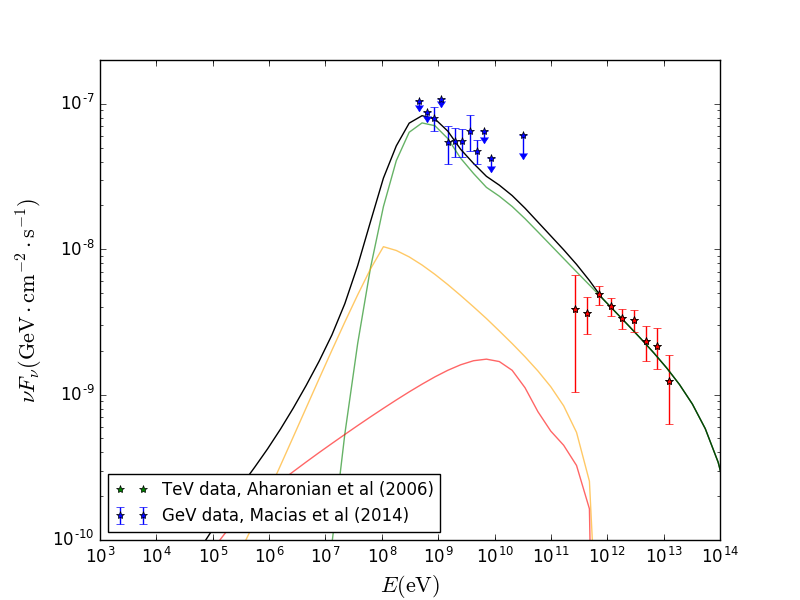}}
\caption{Spectral energy distribution (SED) of the Galactic Ridge for a steady state scenario. The CR escape is due to the advection of the perpendicular wind of speed 1000 km/s on a 30 pc scale on the top panel and to the diffusion on the bottom panel. We assume a power-law diffusion coefficient: D=$D_0 {\left(E/10 \ TeV \right)}^{d}$ with $\rm D_0$=$2 \times 10^{29}$ $\rm cm^2s^{-1}$ and d=0.3. The spectral index of the parent particle population is fixed to 2.45 in the top panel and 2.15 in the bottom panel. The SN recurrence time is fixed around ${10}^4$ years in the top panel and $2 \times 10^3$ years in the bottom panel. For the IC, we considered two target photon populations: an optical radiation field (T=5000K) with an energy density of 60 $\rm{eV \, cm^{-3}}$ and far infrared radiation field (T=21 K) with an energy density of 15 $\rm{eV \, cm^{-3}}$. We assume an efficiency for CR acceleration of 10\% of the kinetic energy released from a SN explosion, $E_k$=$10^{51}$ erg, and a ratio electrons/protons equal to 1\%. The interstellar medium density is equal to 100 $ \rm cm^{-3}$ and the magnetic field to 30 $\rm{\mu G}$.}
\label{SED}
\end{figure}

For the advection case, we assume a power-law for the parent particles population, protons and electrons, with a spectral index $p$ of 2.45. The SED is represented on Figure \ref{SED} (top panel) with the GeV data points from \citet{2014PhRvD..89f3515M} and the TeV data from \citet{2006Natur.439..695A}. By assuming a magnetic field arnd 30 $\rm{\mu G}$, intermediate value of the magnetic field estimations in the general ISM \citep{2011MNRAS.413..763C, 2005ApJ...626L..23L}, we ensure that the model matches with the $\gamma$-ray spectrum observed at TeV energies. To do so, as shown above, we have to consider a quite large SN recurrence time around $10^4$ years.

The figure \ref{SED} (bottom panel) represents the SED for a CR escape due to the diffusion with a power-law diffusion coefficient: $D=D_0 {\left(E/10 \ \rm {TeV} \right)}^{\rm d}$ with $D_0=2 \times 10^{29}$ $\rm cm^2s^{-1}$ and $d$ = 0.3. For diffusive escape, the injection spectrum is harder with a spectral index of 2.15 increasing the IC and bremsstrahlung components of the total SED. As emphasized in the section \ref{diffusion}, the diffusion model matches well with the $\gamma$-ray spectrum observed at TeV energies by assuming a smaller SN recurrence time of $2 \times 10^3$ yrs closer to some GC estimates \citep[e.g.][]{2011MNRAS.413..763C, 2015MNRAS.453..172P}.

For the diffusive or advective scenario, we do not attempt to explain the synchrotron emission because of the multiple phenomena to be taken into account (self-absorption, thermal emission etc) and because the magnetic field is likely highly non uniform making the box model very unlikely. We simply ensure we do not over predict the radio data points from \citet{2011MNRAS.413..763C}.


Considering coefficient diffusion consistent with values deduced from local observations, the diffusive escape is much more dominant over an advective escape at these VHE energies. While the advective scenario requires SN rates in the lower ranges of  estimates, the diffusive scenario requires larger rates well within the range. However, for the typical estimate of the SN rate found though the diffusive approach, the SN recurrence time is larger than the escape time, even for the low energy part of the proton spectrum. Therefore, a stationary state is not reached and one cannot use this approximation for modelling the SNRs in the GC. In the following sections, we will consider a non steady state approach. We investigate the impact of the sources distribution on the $\gamma$-ray emission morphology and CR profile with a non uniform source distribution in a 3D non steady state model.


\section{A simple time dependent 3D model of CR injection and gamma-ray production}
\label{model}
In this section, we describe the 3D distribution of CR impulsive accelerators and gas that we build to explore the impact of the actual matter and SN spatial distribution on the resulting $\gamma$-ray emission.

\subsection{Distribution of CR accelerators}
\label{CR_accelerators}
As mentioned in the section \ref{intro}, one of the key point for studying the CR distribution is the temporal and spatial distribution of the impulsive CR accelerators filling the GC. 

Up to now, more than 100 Wolf-Rayet (WR) stars and O supergiants have been spectroscopically identified in the Galactic Center region \citep{2010ApJ...725..188M}. These stars are spread throughout the CMZ probably due to high isolated massive star formation occurring in the region \citep{2010ApJ...725..188M}. Therefore a significant fraction of SNe has to be broadly distributed over the CMZ. 

A fraction of these massive stars are located in the known cluster members: the Quintuplet, 3-5 Myrs \citep{2004ApJ...611L.105N}, the Arches, 2-3 Myrs \citep{1999ApJ...525..750F} and the central disk surrounding SgrA$^\star$, 4-6 Myrs \citep{2013ApJ...770...44L}. SNRs and PWNs have been observed in the Quintuplet \citep{2013MNRAS.428.3462H, 2015MNRAS.453..172P}. Two relatively young neutron stars have been recently detected in the Central disk: the one likely powering the PWN candidate G359.95-0.04 close to SgrA$^\star$ \citep{2006MNRAS.367..937W} and the magnetar SGR J1745-2900 \citep{2013ApJ...770L..23M}. Considering the age of the last two clusters and the detection of sources resulting from massive star explosions, recent violent phenomena capable of injecting high energy CR such as SN have already occurred in these clusters. The spatial distribution is therefore non uniform. 

In order to model correctly the SN rate in the various spatial components, we have to estimate the SN rate in the clusters. With an estimated age of 2-3 Myrs \citep{1999ApJ...525..750F}, the Arches cluster has probably not experienced any supernova. We therefore do not include a concentration of SNe at this position. The central part, $r \lesssim 0.5 \rm{pc}$, of both the Quintuplet and the Central disk present a top heavy IMF with a slope $\alpha \sim 1.7$ \citep{2013ApJ...764..155L, 2012A&A...540A..57H} much flatter that the traditional Salpeter one $\alpha = 2.35$ \citep{2001ASPC..228..187K}. The total masses estimated in these central regions are respectively $\sim 6\times10^3 \, \rm{M_{\odot}}$ above $0.5 \, \rm{M_{\odot}}$ for the Quintuplet \citep{2012A&A...540A..57H} and $\sim 14-37\times10^3 \, \rm{M_{\odot}}$ above $1 \, \rm{M_{\odot}}$ for the Central disk \citep{2013ApJ...764..155L}. Using the Starburst99 code \citep{1999ApJS..123....3L, 2005ApJ...621..695V} with the previous IMF and total mass estimations, we can estimate a SN rate in the central part of these clusters. We find $0.3 \times 10^{-4}$ for the Quintuplet and $0.3-1.2 \times 10^{-4} \, \mathrm {yrs}^{-1}$ for the central disk. For the outer parts of these clusters, since the mass segregation that will occur at different radii will change the IMF, the extrapolation on the SN rate is not linear and therefore uncertain. 

To model the CR injected by these individual SNRs spread in the CMZ, we generate impulsive sources distributed according to a Poisson law of recurrence time $\tau$=2500 years based on the central value estimated by \citet{2011MNRAS.413..763C}. The spatial distribution is composed of two components. A component where the SNs are uniformly distributed in a cylinder of 150 pc radius and 10 pc height. In addition to the uniform distribution, we assume a concentration of the SNe in the compact and massive star clusters. The SN rate in both clusters ($8 \times 10^{-5} \rm{yrs}^{-1}$) is compatible with the one previously determined from physical observations.

\subsection{Particle propagation}
We model the CR propagation in the GC assuming an isotropic diffusion. The global CR propagation in the interstellar medium is then described by the simplified transport equation \citep{1990acr..book.....B}:

\begin{center}
\begin{eqnarray}
\dfrac{\partial f}{\partial t} +(\vec{u} \cdot \vec{\nabla})f +  D\Delta{f} = Q
\label{transport1}
\end{eqnarray}
\end{center} 
where Q, D$\Delta{f}$ and ($\vec{u} \cdot \vec{\nabla}$)f describe respectively the CR injection, the spatial diffusion and the convection. 

We consider a power-law diffusion coefficient (section \ref{diffusion}) and we assume the typical $d=0.3$. As discussed in section \ref{diffusion}, a high diffusion coefficient is likely and we assume here a value of $2\times 10^{29}$ $\rm cm^2s^{-1}$ at 10 TeV, making advection negligible in the considered energy range. The injection is assumed to be point-like both in time and space. For simplicity, we neglect acceleration of electrons since their contribution should probably be negligible. The 3D Green function, obtained assuming a CR density equal to zero at $r=+\infty$, gives the solution for an impulsive accelerator as the SNRs. We also investigate the scenario of a single accelerator at the GC to follow the interpretation of \citet{2016Natur.531..476H} of the CR profile inferred from the $\gamma$-ray data. Integrating the previous Green function gives the solution for a continuous source \citep{1990acr..book.....B}. 

We assume a power-law CR injection: $N_o E^{-a}\delta(r-r_0)\delta(t_0)$ for the impulsive injection and $N_o \delta(r-r_0) E^{-a}H(t_0)$ for the stationary source with a spectral index $a$ equal to 2 since this value is typical for diffusive shock acceleration. Time dependent CR escape from the SNR shock is not taken into account. SNe are modelled by impulsive CR injections all emitted at the same time whatever their energies.  

There is a limit to the isotropic diffusion for small distances around the source. The mean free path, $r\rm_{mfp}=3D/v$, of the random walk characterizing the particle diffusion along the magnetic field lines can reach 100 pc for the highest CR energy simulated (1 PeV). For this energy, the mean free path is of the order of magnitude of the box size. Therefore the diffusion approximation for this distance is wrong and we need a better description of CR propagation and of the $\gamma$-ray flux they produced in a ballistic regime where particles are barely scattered and their distribution is not isotropized. Here we adopt a simple modification of the diffusion equation solution on short distances: we fix a constant value of the CR density for the distances $r < r_o=3 \times r_{mfp}$ equal to the CR density at $r=r_o$. This approach neglects the $\gamma$-ray emission from the non-isotropized CRs. The contribution of the latter should appear as a weakly extended source. This simple approach allows to better take into account the diffuse $\gamma$-ray emission in the vicinity of a stationary accelerator.

\subsection{Matter distribution}
\label{Matter}

As mentioned in the section \ref{intro}, the CMZ contains around 10\% of the total molecular mass of the Galaxy in a rather complex setup of dense molecular clouds. Around 40\% of the matter in the molecular clouds are spread in four main complexes: Sgr A, Sgr B, Sgr C and the 1.3 complex (located to the Galactic east from Sgr B). An additional widespread high-temperature and lower density diffuse molecular component has also been observed \citep{1998A&A...331..959D}. 

The molecular line like CS or CO are the main tracers of the dense cores of molecular gas. However, since our knowledge of the true gas kinematics in the GC is very limited, the methods used to convert the measured line of sight (LOS) velocity into a radial distance are unreliable. In order to model the 3D $\gamma$-ray distribution produced by the interaction of the CRs with the matter in the GC, one has to build a coherent 3D matter distribution. We will rely on the work of \citet{2004MNRAS.349.1167S} and \citet{2007A&A...467..611F}.

\citet{2007A&A...467..611F} built a 3D matter model, valid for the innermost 3.0 kpc of our Galaxy, for the different gas components (molecular, atomic and ionized) that best fit the observational data and is consistent with the theoretical predictions. They provide, an analytical expression for the elliptical shape of the CMZ using the work of \citet{2004MNRAS.349.1167S} as the first building block of their model, in particular for the center of the ellipse. \citet{2004MNRAS.349.1167S} derived a face-on map of the molecular gas without any kinematics assumption at a Galactic latitude b equal to 0$^\circ$. Their work rely only on the comparison of molecular line surveys: the CO 2.6-mm emission with the OH 18-cm absorption. The GC is itself an intense diffuse 18-cm continuum region. Therefore, the OH line traces preferentially the matter in front of the GC where the OH absorption will preferentially arise whereas the CO line traces the molecular gas equally in the whole CMZ. They assume an axi-symmetric distribution for the OH continuum emission. 

\begin{figure}
\centering
{\includegraphics[width=\columnwidth]{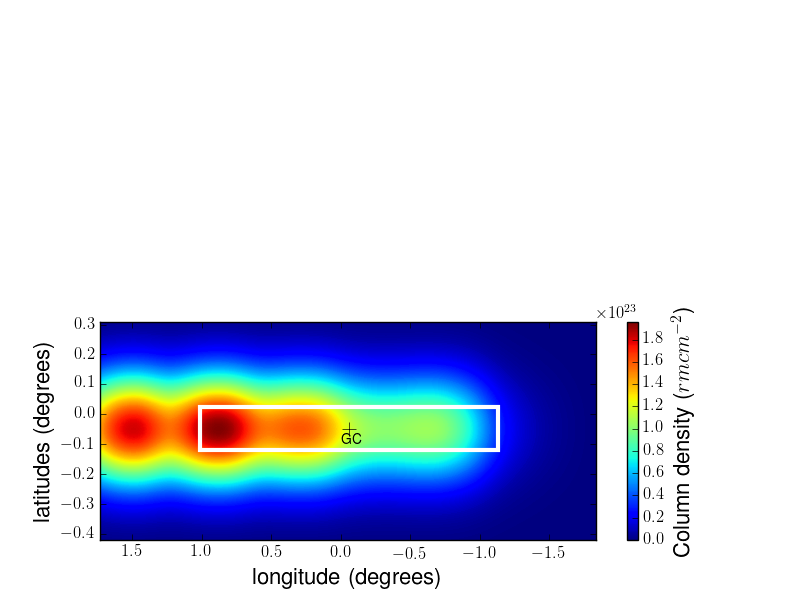}}
{\includegraphics[width=\columnwidth]{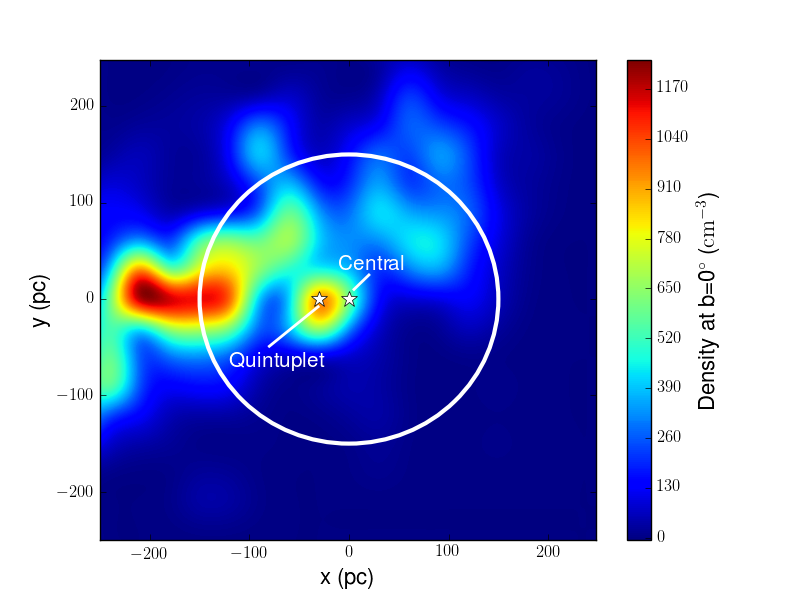}}
\caption{(Top) The resultant molecular face-on view of the Galactic Centre  and (bottom) The resultant molecular top view of the Galactic Centre at the Galactic latitude b = 0$^\circ$ seen from the direction of the north Galactic pole. The region where we draw the spatial distribution of the SNs: a cylinder of 150 pc radius and 10 pc height for the uniform component (white circle on the low panel and white box on the top panel), as well as the position of the Central Cluster and the Quintuplet where we assume a concentration of these sources.}
\label{sawada}
\end{figure}

Assuming an intermediate total mass value of $4\times 10^7$ M$\rm_{\odot}$, we can convert their face-on view intensity map in a density map assuming a same scaling factor in the whole CMZ. As no information on the $\rm H_2$ vertical gas distribution is provided, following \citet{2007A&A...467..611F}, we assume a gaussian decay along the Galactic latitude (top panel of the figure \ref{sawada} ). For the atomic gas, we used a HI mass around 10\% of the H2 mass in the CMZ using the spatial distribution in \citet{2007A&A...467..611F}. The face-on view of our model is represented on the figure \ref{sawada} (top panel) and the top view of our model at a latitude b = 0$^\circ$ seen from the direction of the north Galactic pole on the figure \ref{sawada} (bottom panel). The matter is more spread along the line of sight at larger longitudes.

\subsection{3D model: setup and physical parameters}

We simulate the injection and propagation of CRs from 1 TeV to 1 PeV in a 3D box of size $500 \, \rm {pc} \times 500\, \rm {pc} \times 50 \, \rm {pc}$ centered on the GC. For the impulsive CR accelerators temporal distribution, we don't consider sources with an age $<1 \, \rm{kyr}$ since no younger SNR has been observed within the GC. We only keep sources with an age $<$ 100 kyrs since for older events the CR density becomes negligible. The table \ref{model_param} contains the different physical parameters used in our model. The total $\gamma$-ray flux produced by the pp interaction is obtained by integrating the analytic shape given by \citet{2006PhRvD..74c4018K} for this interaction with the CR spectrum and the 3D matter distribution (section \ref{Matter}) over all the energies of the incident CRs. 

\begin{table}
\caption{Physical parameters values used in our 3D model of the CR injection and propagation.}
\label{model_param}
\begin{tabular}{|p{4cm}|p{4cm}|}
	\hline
   Model Parameters & Values\\
   \hline
   Proton spectral index & 2 \\
   $E_{max}$ of the injected proton spectrum & 1 PeV \\
   Box size & $500 \, \mathrm pc \times 500\, \mathrm pc \times 50 \, \mathrm pc$ \\
   Total gas mass  & $4\times 10^7$ $\rm{M_{\odot}}$  \\
   $D_0$ (10 TeV) & $2\times 10^{29}$ $\rm cm^2s^{-1}$ \\
   Spectral index of the diffusion coefficient (d) & 0.3 \\
   $E_{SN}$ & $10^{51}$ $\rm erg$ \\
   SN recurrence time & 2500 yrs \\  
\hline 
\end{tabular}
\end{table}

\section{Results and discussion}
\label{results}

In this section we compare our model predictions to HESS data, both spectrally and morphologically. To do so we compute the total $\gamma$-ray spectrum predicted in the region as well as the expected $\gamma$-ray and reconstructed CR 1D profiles.

%
\subsection{$\gamma$-rays: spectral distribution}
\label{spectral_distribution}
\begin{figure}
\centering
\subfigure{\includegraphics[width=\columnwidth]{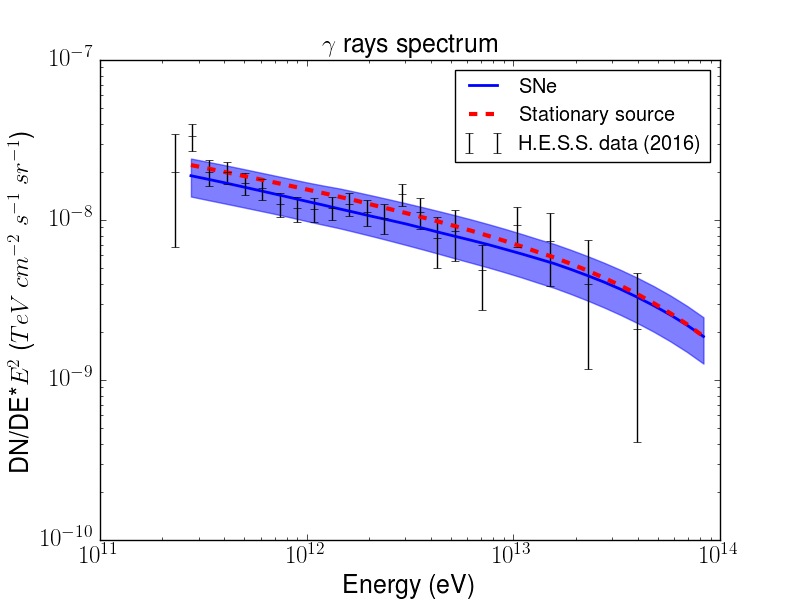}}
\caption{Median of the spectrum generated from 100 SN temporal and spatial distributions (blue line) as well as the dispersion around the median and the spectrum from a stationary source at the GC (red). Both are extracted from an annulus centred at SgrA$^\star$ with inner and outer radii of 0.15$^\circ$ and 0.45$^\circ$. HESS flux points from \citet{2016Natur.531..476H} are shown in black.}
\label{spectrum}
\end{figure}

On Figure \ref{spectrum}, we compare the $\gamma$-ray spectrum in an annulus centred on SgrA$^\star$ with inner and outer radii of 0.15$^\circ$ and 0.45$^\circ$ for a stationary source at the GC and for multiple impulsive injections throughout the GC. For the latter, we generate 100 temporal and spatial distributions of CR accelerators. The spectrum is the median of the previous one hundred Monte Carlo (MC) realizations as well as the dispersion around this median. Using reasonable parameters on the SN rate and the particle diffusion, both our models reproduce the total spectrum observed with H.E.S.S. in the region \citep{2016Natur.531..476H}. 

For the stationary source, we have to assume a power for CR acceleration around $10^{38}$ $\rm erg \cdot s^{-1}$. This power is of the same order of magnitude than the one found by \citet{2016Natur.531..476H} considering that the $\gamma$-ray emission of the central source HESS J1745-290 is due to CRs in a ballistic regime interacting with the matter in the central cavity. Moreover, it represents 10\% of the Bondi accretion power \citep{2013Sci...341..981W} so this value seems plausible. For the SNe, we have to consider a relatively low acceleration efficiency of 2\% of the kinetic energy released from a SN explosion, $E_k$=$10^{51}$ erg. Considering a higher efficiency close to 10\%, as commonly used for SN acceleration, the resulting $\gamma$-ray flux from the SNe is larger than the observations. It is possible to decrease this flux by adjusting the values of other parameters of the model. We can lower the SN rate compare to the central one estimated in \citet{2011MNRAS.413..763C} but it will be in the lower range of the estimated SN rates. We can consider a lower molecular mass for the CMZ at the lower boundaries of the different total mass estimations.

Exploring the variances of different input parameters of the model as the SN rate ($10^{-4}-10^{-3} \, \rm{yrs^{-1}}$) or the total mass of the CMZ (1.2-6.4$\, \times 10^7$ M$\rm_{\odot}$), in order to reproduce the total $\gamma$-ray flux, the efficiency for the CR acceleration goes from 0.5\% to 20\%. Any larger efficiency value will overproduce the flux. The upper edge is still in the order of magnitude of the common acceleration efficiency expected for the SNRs. The lower edge seems really low, it's difficult to explain why the CR acceleration in the SNRs will be so inefficient. Even by taking into account the incertitudes of some input parameters, the SN contribution is a significant fraction of the total VHE flux. For certain values of the input parameters, if we do not consider extremely low CR acceleration efficiency values in the SNR, they already overproduce the total flux.

\subsection{Morphology emission}
\label{spatial_distribution}
\subsubsection{$\gamma$-ray profile}
\label{gamma_profile}
\begin{figure}
\centering
{\includegraphics[width=\columnwidth]{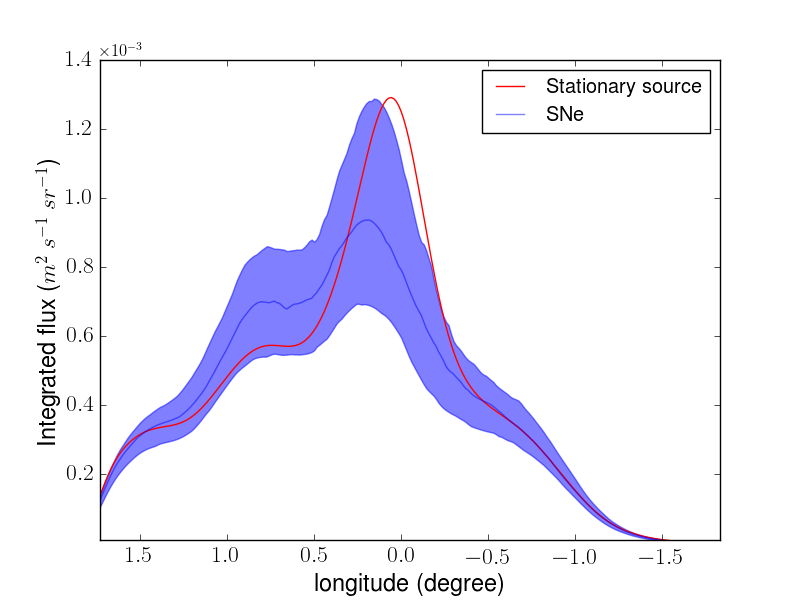}}
{\includegraphics[width=\columnwidth]{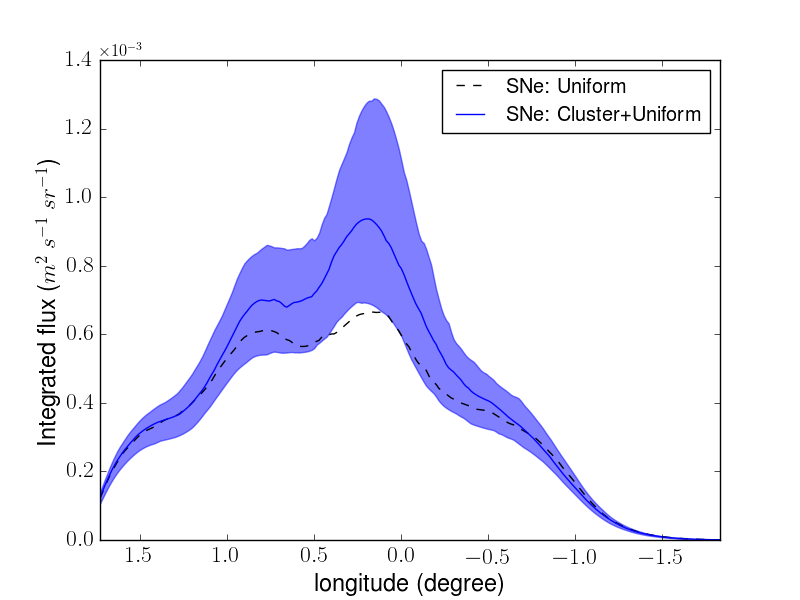}}
\caption{(Top) Expected VHE $\gamma$-ray profile along the Galactic longitude for a distribution of the SNe taking into account the two massive clusters (blue) and for a stationary source at the GC (red), after integrating along the line of sight and the Galactic latitude b. (Bottom) Expected VHE $\gamma$-ray profile along the Galactic longitude produced by the SNe throughout the CMZ for two spatial distributions: a unique uniform distribution of the SNe in the GC (black dashed line) and a distribution of the SNe taking into account the two massive clusters: the Quintuplet and the Central disk in addition to the uniform distribution (blue). The solid and dashed lines represent the median of the 100 spatial and temporal SN distributions and the colored regions the dispersion around this median.}
\label{profile}
\end{figure}

The 1D profile along the Galactic longitude resulting from a stationary source located at the GC obtained by integrating the predicted emission for latitudes $\rvert b \rvert <0.3^\circ$ is represented in red in Figure \ref{profile} (top panel). Considering the deficit of VHE emission beyond 100 pc (in particular at l=1.3$^\circ$) relative to the available target material, \citet{2006Natur.439..695A} concluded to an impulsive injection at the center 10 kyrs ago. Here, the $\gamma$-ray profile drops at $\rvert l \rvert >1.3^\circ$ as the result of the integration of CR density with a more spread matter distribution along the line of sight as observed in the Figure \ref{sawada} (bottom panel). 

Figure \ref{profile} (bottom panel) represents the median of the 1D profile of the 100 MC realizations as well as the dispersion around this median for the SNe distributions. Taking into account a realistic spatial distribution for the SNe with a concentration in the central clusters in addition to the uniform distribution (blue curve), one finds a $\gamma$-ray profile which peaks towards the GC. As indicated by the huge dispersion around the median, the profile strongly depends on the realization of both spatial and temporal distributions of the CR accelerators since the more recent explosions have a more significant impact on the resulting $\gamma$-ray profile. As expected, the median of the profiles predicted if we consider only the uniform component of the accelerators in the spatial distribution is rather flat and peaked on the well-known molecular clouds in the GC (black dashed line on the bottom panel of Figure \ref{profile}). 

The $\gamma$-ray profiles obtained from either a stationary source or from a realistic SN spatial distribution (Figure \ref{profile}, top panel) are very similar and an increase of the $\gamma$-ray emission towards the center, as detected by \citet{2016Natur.531..476H}, is obtained in both models. However, the profile produced by a stationary source is more peaked towards the source itself, here assumed to be at the GC, than the one produced by the multiple accelerators which is peaked on the molecular clouds next to SgrA$^\star$. We then explore this difference comparing the expected CR profiles to the one deduced from HESS observations. 

\subsubsection{CR density profile}

\begin{figure}
\centering
\subfigure[]{\includegraphics[width=\columnwidth]{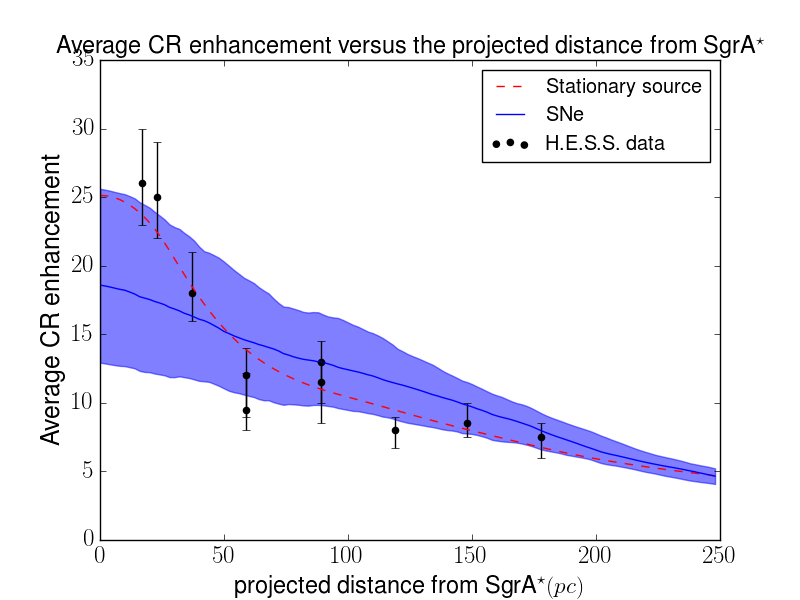}}
\caption{Average CR enhancement with distance to the GC (see text) from \citet{2016Natur.531..476H} in black, for a stationary source situated at the GC in red and for the SNe model taking into account the two massive clusters for the spatial distribution in blue. These profiles are the mean of the profiles for Galactic latitudes: $\rvert b \rvert <0.1^\circ$.}
\label{CRenhancement}
\end{figure}

Using the work of \citet{2016Natur.531..476H}, it is possible to directly compare the CR density profile derived from our models with the one obtained from the H.E.S.S. data. They extracted the VHE $\gamma$-ray luminosity within seven circular regions of radius 0.1 degrees distributed along the Galactic Plane. For each of these regions, they deduced the average CR density integrated over the line of sight as a function of the projected distance from SgrA$^\star$ (at 8.5 kpc). Within each region, the gas mass, $M_{gas}$, is estimated from the CS emission line. They deduced the energy density of CR above 10 TeV (those producing VHE $\gamma$-rays above 1 TeV) needed to explain the $\gamma$-ray luminosities $L_{\gamma}$ and scaled it as $L_{\gamma}$/$M_{gas}$ in each of these regions. Figure \ref{CRenhancement} presents the average CR enhancement within these regions compared with the local CR energy density measured in the Solar neighbourhood at those energies which is $w_0(> 10 \, \rm{TeV}) \approx 10^{-3}$ eV/cm3.

In order to compare our models with these results, we determine the average energy density of CR above 10 TeV along the LOS by weighting the $\gamma$-ray luminosity by our matter distribution:
\begin{center}
\begin{eqnarray}
n_{CR}(x,z) = \frac{\int_{y} L_{\gamma}(x,y,z) \times n(x,y,z) \, dy }{\int_{y} n(x,y,z) \, dy }
\end{eqnarray}
\end{center}
where $n_{CR}$ is the CR density in Galactic latitude ($z$) and Galactic longitude ($x$), $n(x,y,z)$ the matter density in each pixel of the 3D box, $L_{\gamma}(x,y,z)$ the $\gamma$-ray luminosity and $y$ the direction along the LOS.

With this approach, the CR density profiles for the SNe model and the stationary source presented in Figure \ref{CRenhancement} are independent of the total molecular mass assumed in our model. This is a significant difference with the H.E.S.S. data points extracted in \citet{2016Natur.531..476H}, where the absolute error on the conversion of CS line into $\mathrm H_2$ column density has to be taken into account. This is why we have to consider a correction factor between our models and the data from \citet{2016Natur.531..476H} due to the uncertainty on the total mass.

Taking into account a realistic SNe and matter spatial distributions provides a peaked $\gamma$-ray profile (section \ref{gamma_profile}) and a pronounced CR gradient as seen in Figure \ref{CRenhancement}. The observed CR profile seems more peaked than the profile predicted by SNe alone. A second VHE $\gamma$-ray component is probably required in the central 30 pc in addition to the SN contribution. Although a GC stationary source alone can well explain the CR profile as already noted by \citet{2016Natur.531..476H}, it requires negligible contribution from SNe which is at odds with the already low acceleration efficiency required to not over-produce the total $\gamma$-ray spectrum (section \ref{spectral_distribution}). However acceleration efficiency might be significantly different than what it is in the rest of the Galaxy.
First, if most SNe take place in the hot (kT$\sim$1 keV) and relatively dense phase ($\sim$1 $\rm{cm^3}$) of the GC, shock waves should become mildly supersonic in a short amount of time \citep{2005ApJ...628..205T}. Regular efficient diffusive acceleration could therefore occur over a limited time period. 
Second, many SNe occur in groups and although conditions are somewhat different from typical superbubbles (e.g. because of the Quintuplet cluster motion), collective effects should also play a role \citep{2014A&ARv..22...77B}.

\section{Conclusions}
We studied the contribution of CRs accelerated by SNRs spread throughout the GC on the VHE emission observed with H.E.S.S. in the inner 200 pc of the Galaxy. The input power of these multiple accelerators in the GC is much higher than the $\gamma$-ray luminosity observed with H.E.S.S.. Therefore an efficient CR escape from the region is necessary regarding this high input power. Using a stationary steady state model, we studied how the CRs escape from the GC and we concluded that at these very high energies a diffusive escape is dominant over an advective escape given typical diffusion coefficient value in the ISM.

In order to study the impact of the CR accelerators spatial distribution, we build a 3D model of CR injection and diffusion with a realistic 3D gas distribution. We found that, assuming a characteristic value of 2500 yrs for the recurrence time of these events, the 3D models of random burst like injection of CRs following SN time and spatial distribution throughout the CMZ reproduce the total $\gamma$-ray flux observed with H.E.S.S. in the GC. We have to consider a relatively low acceleration efficiency of 2\% of the kinetic energy released from a SN explosion. The 3D spatial distribution of these accelerators has a major impact on the morphology of the CR density and on the resulting $\gamma$-ray VHE emission. Taking a realistic spatial distribution for the SNe composed of a uniform component and a concentration of theses sources in the massive star clusters of the GC, the Quintuplet and the Central disk, creates a pronounced CR gradient and makes the VHE $\gamma$-ray and CR profile peaked towards the GC. At larger longitudes, the CR density produced by this model fits the profile recently obtained by HESS. For distances inferior to 30 pc from the GC, a second VHE component is required to reproduce the very central excess on top of the SNe contribution.

We also tested the scenario proposed by \citet{2016Natur.531..476H} in which a single scenario of a stationary source at the center models the central source SgrA$^\star$ since the CR density profile they extracted from H.E.S.S. data is compatible with a 1/r profile. We found a power for CR acceleration around $10^{38}$ $\rm erg \cdot s^{-1}$ i.e. 10\% of the Bondi accretion power as found by \citet{2016Natur.531..476H} making this central source a good candidate to the VHE CR acceleration in the GC. 

However, even if the excess of CR and VHE $\gamma$-ray emission observed towards the center is well reproduced by this stationary source scenario alone, it is too simplistic to conclude that such a source is the only source responsible for all the VHE emission observed with H.E.S.S. in the GC. We have to take into account the recurrent CR injections from the numerous SNe observed through the CMZ that explain already the data at distances larger than 30 pc. Considering a scenario with only SgrA$^\star$ would imply to reduce the SNR contributions of CRs to very low efficiencies. This possibility should not be excluded considering the very hot medium where the SNe are located. It is possible to consider weakly supersonic shocks to decrease the SN efficiency. We could also consider a higher recurrence time for the SNe but this seems at odds with various observations, the SN rate estimated only in the central clusters beeing already high.

Taking into account the morphology of the magnetic field observed in the GC, which is approximately poloidal on average in the diffuse intercloud medium \citep{2014arXiv1406.7859M}, the diffusion coefficient is probably anisotropic. Knowing that we are using a conservative value for the isotropic diffusion coefficient, we could consider an anisotropic diffusion coefficient with a higher value in the direction perpendicular to the Galactic plane: this would mechanically decrease the contribution of SNe to the VHE signal, except if we increase the CR acceleration efficiency. Even by increasing this diffusion coefficient by a factor 4, the model is still compatible with the H.E.S.S. data by taking an acceleration efficiency closer to 10\%.

\section*{Acknowledgements}
We acknowledge the Fran\c cois Arago Centre at the APC laboratory to provide the resources necessary to compute the simulations presented in this work. We also acknowledge the whole team that developed the software Starburst99 used to determine the SN rate in the central Clusters of the Galaxy. We thank kindly Joachim Hahn for the implementations he made very quickly for us in the software GAMERA. We thank Maica Clavel and Stefano Gabici for the useful discussions and the careful review of this manuscript.
%
%
%
%
%
\bibliographystyle{mnras}
\bibliography{biblio} 



%
%


\bsp	
\label{lastpage}
\end{document}